\documentclass[%
 reprint,
 amsmath,amssymb,
 aps,
]{revtex4-2}

\usepackage[dvips]{graphicx}
\usepackage{dcolumn}
\usepackage{bm}
\usepackage{braket}

\def\Journal#1#2#3#4{{#1} {\bf #2}, #3 (#4)}
 
\def\AandA{Astron. \& Astrophys.}

\def\CPC{Chin. Phys. C}
\def\EPJC{Eur. Phys. J. C}
\def\EPL{EPL}

\def\JHEP{JHEP}

\def\JPG{J. Phys. G} 
\def\MPLA{Mod. Phys. Lett. A}

\def\NPB{Nucl. Phys. B}

\def\PL{Phys. Lett.}
\def\PLB{{Phys. Lett.} B}

\def\PRL{Phys. Rev. Lett.}
\def\PRD{Phys. Rev. D}

\def\PTP{Prog. Theor. Phys.}
\def\PTEP{Prog. Theor. Exp. Phys.}

\def\SCIENCE{Science}

\def\ZPC{Z. Phys. C}

\begin{document}


 \title{Texture zeros flavor neutrino mass matrix and triplet Higgs models}

\author{Teruyuki Kitabayashi}
\email{teruyuki@tokai-u.jp}

\affiliation{%
\sl Department of Physics, Tokai University, 4-1-1 Kitakaname, Hiratsuka, Kanagawa 259-1292, Japan
}


\begin{abstract}
One- and two-zero textures for the flavor neutrino mass matrix have been successful in explaining mixing in the neutrino sector. Conservatively, six cases of one-zero textures and seven cases of two-zero textures are compatible with observations. We show that one case may be the most natural in the one- and two-zero textures schemes if tiny neutrino masses are generated by the type-II seesaw mechanism in triplet Higgs models.
\end{abstract}

\maketitle


\section{Introduction\label{sec:introduction}}
The origin of the tiny masses and flavor mixing of neutrinos is a long-term mystery in particle physics. The seesaw mechanism is one of the leading theoretical mechanisms for generating tiny neutrino masses. There are three types of seesaw mechanisms \cite{Zhou2015POS}. 
\begin{enumerate}
\item Type I: Right-handed singlet neutrinos are introduced in the standard model \cite{Minkowski1977PLB,Yanagida1979KEK,Gell-Mann1979,Glashow1979,Mohapatra1980PRL}. 
\item Type II: A triplet scalar (triplet Higgs boson) is introduced in the standard model \cite{Konetschny1977PLB,Schechter1980PRD,Cheng1980PRD,Magg1980PLB,Lazarides1981NPB,Mohapatra1981PRD}.
\item Type III: Triplet fermions are introduced in the standard model \cite{Foot1989ZPC}. 
\end{enumerate}

To solve the origin of flavor mixing of neutrinos, there have been various discussions on the texture zeros approach for flavor neutrino masses \cite{Ludl2014JHEP}. In this approach, we assume that the flavor neutrino mass matrix has zero elements. 

In the one-zero texture scheme, there are the following six cases for the flavor neutrino mass matrix:
\begin{eqnarray}
&& {\rm G}_1:
\left( 
\begin{array}{*{20}{c}}
0 & \times & \times \\
- & \times & \times \\
- & - & \times \\
\end{array}
\right),
\quad
 {\rm G}_2:
\left( 
\begin{array}{*{20}{c}}
\times & 0 & \times \\
- & \times & \times \\
- & - & \times \\
\end{array}
\right),
\nonumber \\
&&{\rm G}_3:
\left( 
\begin{array}{*{20}{c}}
\times & \times &0 \\
- & \times & \times \\
- & - & \times \\
\end{array}
\right),
\quad
{\rm G}_4:
\left( 
\begin{array}{*{20}{c}}
\times & \times & \times \\
- & 0 & \times \\
- & - & \times \\
\end{array}
\right),
\nonumber \\
&& {\rm G}_5:
\left( 
\begin{array}{*{20}{c}}
\times & \times & \times \\
- & \times & 0 \\
- & - & \times \\
\end{array}
\right),
\quad 
{\rm G}_6:
\left( 
\begin{array}{*{20}{c}}
\times & \times & \times \\
- & \times & \times \\
- & - &0 \\
\end{array}
\right).
\label{Eq:G1G2G3G4G5G6}
\end{eqnarray}
All six cases of one-zero textures are consistent with observations \cite{Bora2017PRD}. 

In the two-zero texture scheme, there are 15 possible combinations of two vanishing independent elements in the $3 \times 3$ Majorana flavor neutrino mass matrix. The neutrino oscillation data allows only 7 out of the 15 cases  \cite{Fritzsch2011JHEP,Dev2014PRD,Meloni2014PRD,Zhou2016CPC,Singh2016PTEP},
\begin{eqnarray}
&& {\rm A}_1:
\left( 
\begin{array}{*{20}{c}}
0 & 0 & \times  \\
- & \times  & \times \\
- & - & \times \\
\end{array}
\right),
\quad
 {\rm A}_2:
\left( 
\begin{array}{*{20}{c}}
0& \times  &0 \\
- & \times & \times \\
- & - & \times  \\
\end{array}
\right),
\nonumber \\
&& {\rm B}_1:
\left( 
\begin{array}{*{20}{c}}
\times & \times & 0 \\
- & 0 & \times \\
- & - & \times \\
\end{array}
\right),
\quad
 {\rm B}_2:
\left( 
\begin{array}{*{20}{c}}
\times & 0 & \times \\
- & \times & \times \\
- & - & 0 \\
\end{array}
\right),
\nonumber \\
&&{\rm B}_3:
\left( 
\begin{array}{*{20}{c}}
\times & 0 &\times \\
- & 0 & \times \\
- & - & \times \\
\end{array}
\right),
\quad
{\rm B}_4:
\left( 
\begin{array}{*{20}{c}}
\times & \times & 0 \\
- & \times & \times \\
- & - & 0 \\
\end{array}
\right),
\nonumber \\
&& {\rm C}:
\left( 
\begin{array}{*{20}{c}}
\times & \times & \times \\
- & 0 & \times\\
- & - & 0 \\
\end{array}
\right).
\label{Eq:A1A2B1B2B3B4C}
\end{eqnarray}
If neutrinoless double beta decay is observed in future experiments,  the ${\rm A_1}$ and ${\rm A_2}$ cases should be excluded \cite{GERDA2019Science,Capozzi2020PRD}. Moreover, Singh shows only ${\rm B_2}$ and ${\rm B_4}$ are compatible with recent data at $2\sigma$ \cite{Singh2020EPL}. In this paper, all seven cases of two-zero textures in Eq.(\ref{Eq:A1A2B1B2B3B4C}) are included in our study in a conservative manner. 

The origin of such texture zeros is discussed in Refs.\cite{Berger2001PRD,Low2004PRD,Low2005PRD,Grimus2004EPJC,Xing2009PLB,Dev2011PLB,Araki2012JHEP,Felipe2014NPB,Grimus2005JPG}. The phenomenology of one-zero and two-zero textures is studied in, for example, Refs.\cite{Xing2004PRD,Lashin2012PRD,Deepthi2012EPJC,Gautam2015PRD,Barreiros2019JHEP,Barreiros2020arXiv,Verma2020MPLA} and Refs. \cite{Cebola2015PRD,Frampton2002PLB,Xing2002PLB530,Xing2002PLB539,Kageyama2002PLB,Dev2007PRD,Ludle2012NPB,Kumar2011PRD,Fritzsch2011JHEP,Meloni2013NPB,Meloni2014PRD,Dev2014PRD,Dev2015EPJC,Kitabayashi2016PRD,Barreiros2018PRD,Correia2019PRD}, respectively.

In this paper, we demonstrate that all six cases of one-zero textures (${\rm G_1}, {\rm G_2}, \cdots, {\rm C_6}$) and all seven cases of two-zero textures (${\rm A_1}, {\rm A_2}, {\rm B_1},\cdots, {\rm B_4}, {\rm C}$) are excluded if the following two conditions are satisfied:
\begin{description}
\item[ C1:] Neutrino masses are generated by the type-II seesaw mechanism in triplet Higgs models.
\item[ C2:] The three lepton flavor violating processes $\mu \rightarrow \bar{e}ee$, $\tau \rightarrow \bar{\mu}\mu\mu$ and $\tau \rightarrow \bar{e}ee$ are  all explicitly forbidden either experimentally or theoretically. 
\end{description}
Moreover, we show that the ${\rm G_6}$ case is viable only if the condition {\bf C1} as well as the following conditions is satisfied:

\begin{description}
\item[ C3:] The three lepton flavor violating processes $\mu \rightarrow \bar{e}ee$, $\tau \rightarrow \bar{\mu}\mu\mu$, and $\tau \rightarrow \bar{e}ee$ are  all observed experimentally or undoubtedly are predicted theoretically.
\end{description}
Even if part of these three lepton flavor violating processes is allowed such as ${\rm BR}(\mu \rightarrow \bar{e}ee) \neq 0$, ${\rm BR}(\tau \rightarrow  \bar{\mu}\mu\mu) =  0$, and ${\rm BR}(\tau \rightarrow \bar{e}ee) = 0$, other cases of one- and two-zero textures may be allowed; however, we show that the ${\rm G_6}$ case may be the most natural in the one- and two-zero textures schemes.

We also show that this conclusion becomes more rigid by including other four lepton flavor violating processes, $\tau \rightarrow \bar{\mu} ee$, $\tau \rightarrow \bar{\mu} e \mu$, $\tau \rightarrow \bar{e} \mu \mu$ and $\tau \rightarrow \bar{e} e \mu$ in our discussions.

The paper is organized as follows. In Sec.\ref{sec:TripletHiggs}, we present a brief review of the triplet Higgs model.  In Sec.\ref{sec:zeros}, we show that the ${\rm G_6}$ case may be the most natural in the one- and two-zero textures schemes if the neutrino masses are generated by the type-II seesaw mechanism in triplet Higgs models. Section \ref{sec:summary} is devoted to a summary.

\section{Triplet Higgs model \label{sec:TripletHiggs}}
We assume that neutrino masses are generated by the type-II seesaw mechanism in triplet Higgs models. In triplet Higgs models \cite{Konetschny1977PLB,Schechter1980PRD,Cheng1980PRD,Magg1980PLB,Lazarides1981NPB,Mohapatra1981PRD}, an $SU(2)$ triplet scalar field
\begin{eqnarray}
\Delta = \left( 
\begin{array}{*{20}{c}}
\xi^+/\sqrt{2} & \xi^{++} \\
\xi^0 & -\xi^+/\sqrt{2} 
\end{array}
\right),
\end{eqnarray}
is introduced into the particle contents of the standard model. This triplet of scalar fields yields a Majorana mass of the neutrinos via the following Yukawa interaction:
\begin{eqnarray}
\mathcal{L} = y_{ij}\psi_{iL}^TC i \tau_2\Delta \psi_{jL} + {\rm H.c.},
\end{eqnarray}
where $y_{ij}$ ($i,j=e,\mu,\tau)$ is the $(i,j)$ element of the complex and symmetric Yukawa coupling matrix, $C$ is the charge conjugation, $\tau_2$ is a Pauli matrix, and $\psi_{iL}=(\nu_i, \ell_i)^T_L$ is a standard model left-handed lepton doublet. After $\xi^0$ develops a nonzero vacuum expectation value $v_\Delta = \braket{\xi^0}$, Majorana neutrino masses $M_{ij}$ are generated. 

One of the most important relations in triplet Higgs models is the one-to-one correspondence between the flavor neutrino masses $M_{ij}$ and Yukawa couplings $y_{ij}$ \cite{Ma2001PRL,Kakizaki2003PLB,Akeroyd2009PRD}:
\begin{eqnarray}
y_{ij} = \frac{M_{ij}}{\sqrt{2} v_\Delta}.
\label{Eq:yM}
\end{eqnarray}

These Yukawa matrix elements $y_{ij}$ are also related to lepton flavor violating processes \cite{Ma2001PRL,Kakizaki2003PLB,Akeroyd2009PRD, Dev2018PRD,Dev2018JHEP,Primulando2019JHEP}. For example, the virtual exchange of doubly charged Higgs bosons induces an effective interaction of four charged leptons for $\ell_m \rightarrow \bar{\ell}_i \ell_j \ell_k$ decay at tree level. The branching ratios for the lepton flavor violating decays $\mu \rightarrow \bar{e}ee$ and $\tau \rightarrow \bar{\ell}_i \ell_j \ell_k$ are given by 
\begin{eqnarray}
{\rm BR}(\mu \rightarrow \bar{e}ee) = \frac{|y_{\mu e}|^2 |y_{ee}|^2 }{4G_F^2M_{\pm\pm}^4} {\rm BR}(\mu\rightarrow e \bar{\nu}\nu),
\end{eqnarray}
and
\begin{eqnarray}
{\rm BR}(\tau \rightarrow \bar{\ell}_i \ell_j \ell_k) = \frac{S |y_{\tau i}|^2 |y_{jk}|^2 }{4G_F^2M_{\pm\pm}^4} {\rm BR}(\tau\rightarrow \mu \bar{\nu}\nu),
\end{eqnarray}
where $S=1 (2)$ for $j=k$ ($j\neq k$), $G_F$ is the Fermi coupling constant and $M_{\pm\pm}$ denotes the mass of the doubly charged Higgs bosons \cite{Akeroyd2009PRD}. 

Thanks to the one-to-one correspondence between the flavor neutrino masses and Yukawa couplings, the branching ratios of the lepton flavor violating decay $\ell_m \rightarrow \bar{\ell}_i \ell_j \ell_k$ directly connect with the neutrino flavor masses,
\begin{eqnarray}
{\rm BR}(\mu \rightarrow \bar{e}ee) &\propto& |M_{e\mu}|^2 |M_{ee}|^2, \nonumber \\
{\rm BR}(\tau \rightarrow \bar{\mu} \mu \mu) &\propto& |M_{\mu\tau}|^2 |M_{\mu\mu}|^2, \nonumber \\
{\rm BR}(\tau \rightarrow \bar{e} e e) &\propto&  |M_{e\tau}|^2 |M_{ee}|^2,
\label{Eq:BR1}
\end{eqnarray}
as well as
\begin{eqnarray}
{\rm BR}(\tau \rightarrow \bar{\mu} ee) &\propto& |M_{\mu\tau}|^2 |M_{ee}|^2, \nonumber \\
{\rm BR}(\tau \rightarrow \bar{\mu} e \mu) &\propto& |M_{\mu\tau}|^2 |M_{e\mu}|^2, \nonumber \\
{\rm BR}(\tau \rightarrow \bar{e} \mu \mu) &\propto& |M_{e\tau}|^2 |M_{\mu\mu}|^2, \nonumber \\
{\rm BR}(\tau \rightarrow \bar{e} e \mu) &\propto& |M_{e\tau}|^2 |M_{e\mu}|^2.
\label{Eq:BR2}
\end{eqnarray}
These simple relations between the branching rations and the flavor neutrino mass matrix in  Eqs. (\ref{Eq:BR1}) and (\ref{Eq:BR2}) are useful for testing the availability of the zero texture of the flavor neutrino mass matrix. For example, we can test the availability of a texture which has $M_{ee}=0$ by using a branching ratio which is proportional to  $M_{ee}$ such as ${\rm BR}(\mu \rightarrow \bar{e}ee)\propto |M_{\mu\tau}|^2 |M_{ee}|^2$.

We would like to note again that the origin of these simple relations in Eqs. (\ref{Eq:BR1}) and (\ref{Eq:BR2}) is the one-to-one correspondence between $M_{ij}$ and $y_{ij}$, [Eq. (\ref{Eq:yM})] in the type-II seesaw mechanism. In the type-I and -III seesaw mechanisms, we obtain more complicated correspondences between $M_{ij}$ and $y_{ij}$, such as the Casas-Ibarra parametrization \cite{Casas2001NPB,Ibarra2004PLB} for the type-I seesaw mechanism. This is the reason why we chose the type-II seesaw mechanism, not type I or III, to explain the neutrino mass along with texture zeros.

In the next section, we use these branching ratios of the lepton flavor violating processes to test the availability of the one- and two-zero textures. First, we will include only three branching ratios in Eq.(\ref{Eq:BR1}) in our discussion to show our strategy. Then we include the remaining four branching ratios in Eq.(\ref{Eq:BR2}) in our discussion to complete this paper.

\section{Texture zeros \label{sec:zeros}}
\subsection{${\rm BR}(\mu\rightarrow 3e,\tau\rightarrow 3\mu, \tau\rightarrow 3e)=0$ case }

In this subsection, we assume that the three lepton flavor violating processes $\mu \rightarrow \bar{e}ee$, $\tau \rightarrow \bar{\mu}\mu\mu$ and  $\tau \rightarrow \bar{e}ee$  are all explicitly forbidden either experimentally or theoretically. 

In this case, at least the branching ratio ${\rm BR}(\mu \rightarrow \bar{e}ee)$, as well as $M_{ee}$ and/or $M_{e\mu}$, should vanish. If we require the conditions of $M_{ee}=0$ and/or $M_{e\mu}=0$ for the ${\rm G_3}$ case in the one-zero textures scheme
\begin{eqnarray}
&&{\rm G}_3:
\left( 
\begin{array}{*{20}{c}}
\times & \times &0 \\
- & \times & \times \\
- & - & \times \\
\end{array}
\right),
\end{eqnarray}
the following three flavor neutrino mass matrix are obtained:
\begin{eqnarray}
\left( 
\begin{array}{*{20}{c}}
0 & \times &0 \\
- & \times & \times \\
- & - & \times \\
\end{array}
\right),
\left( 
\begin{array}{*{20}{c}}
\times & 0 &0 \\
- & \times & \times \\
- & - & \times \\
\end{array}
\right),
\left( 
\begin{array}{*{20}{c}}
0 & 0 &0 \\
- & \times & \times \\
- & - & \times \\
\end{array}
\right).
\end{eqnarray}
However,  the one-zero textures assumption is violated in these matrices by an additional vanishing entry. Therefore, the ${\rm G_3}$ case in the one-zero textures scheme should be excluded if the lepton flavor violating process $\mu \rightarrow \bar{e}ee$ is explicitly forbidden. In the same manner, we can exclude the following ${\rm G_4}$, ${\rm G_5}$, and ${\rm G_6}$ cases, 
\begin{eqnarray}
&& {\rm G}_4:
\left( 
\begin{array}{*{20}{c}}
\times & \times & \times \\
- & 0 & \times \\
- & - & \times \\
\end{array}
\right),
\quad
 {\rm G}_5:
\left( 
\begin{array}{*{20}{c}}
\times & \times & \times \\
- & \times & 0 \\
- & - & \times \\
\end{array}
\right),
\nonumber \\ 
&&{\rm G}_6:
\left( 
\begin{array}{*{20}{c}}
\times & \times & \times \\
- & \times & \times \\
- & - &0 \\
\end{array}
\right),
\label{Eq:G3G4G5G6}
\end{eqnarray}
if the lepton flavor violating process $\mu \rightarrow \bar{e}ee$ is explicitly forbidden.  Moreover, the following ${\rm B_1}$, ${\rm B_4}$, and ${\rm C}$ cases in the two-zero textures scheme,
\begin{eqnarray}
&& {\rm B}_1:
\left( 
\begin{array}{*{20}{c}}
\times & \times & 0 \\
- & 0 & \times \\
- & - & \times \\
\end{array}
\right),
\quad
{\rm B}_4:
\left( 
\begin{array}{*{20}{c}}
\times & \times & 0 \\
- & \times & \times \\
- & - & 0 \\
\end{array}
\right),
\nonumber \\
&& {\rm C}:
\left( 
\begin{array}{*{20}{c}}
\times & \times & \times \\
- & 0 & \times\\
- & - & 0 \\
\end{array}
\right),
\label{Eq:B1B4C}
\end{eqnarray}
are also excluded if we require the conditions of  $M_{ee}=0$ and/or $M_{e\mu}=0$ (the two-zero textures assumption should be violated by this requirement).

Consequently, the ${\rm G_3}$,${\rm G_4}$,${\rm G_5}$, and ${\rm G_6}$ cases of one-zero textures and ${\rm B_1}$, ${\rm B_4}$, and ${\rm C}$ cases of two-zero textures should be excluded if the lepton flavor violating process $\mu \rightarrow \bar{e}ee$ is explicitly forbidden.

In addition to the lepton flavor violating process $\mu \rightarrow \bar{e}ee$, we can use other two lepton flavor violating processes $\tau \rightarrow \bar{\mu}\mu\mu$ and $\tau \rightarrow \bar{e}ee$ to test the compatibility of the one- and two-zero textures. Table \ref{tbl:all_zero} shows the compatibility of the cases in the one- and two-zero textures schemes with the vanishing branching ratios ${\rm BR}(\mu \rightarrow \bar{e}ee)=0$, ${\rm BR}(\tau \rightarrow \bar{\mu}\mu\mu)=0$ and ${\rm BR}(\tau \rightarrow \bar{e}ee)=0$. The symbol $\times$ indicates that the corresponding case should be excluded. 

We conclude that all six cases of one-zero textures (${\rm G_1}, {\rm G_2}, \cdots, {\rm C_6}$) and all seven cases of two-zero textures (${\rm A_1}, {\rm A_2}, {\rm B_1},\cdots, {\rm B_4}, {\rm C}$) should be excluded if the neutrino masses are generated by the type-II seesaw mechanism in the triplet Higgs models, and the three lepton flavor violating processes $\mu \rightarrow \bar{e}ee$, $\tau \rightarrow \bar{\mu}\mu\mu$, and $\tau \rightarrow \bar{e}ee$ are all explicitly forbidden.

\begin{table}[t]
\caption{Compatibility of the cases in the one- and two-zero textures scheme with the vanishing branching ratios ${\rm BR}(\mu \rightarrow \bar{e}ee)=0$, ${\rm BR}(\tau \rightarrow \bar{\mu}\mu\mu)=0$, and ${\rm BR}(\tau \rightarrow \bar{e}ee)=0$. The symbol $\times$ indicates that the corresponding case should be excluded. }
\begin{center}
\begin{tabular}{|c|c|c|c|}
\hline
 & ${\rm BR}(\mu \rightarrow 3e) =0$ & ${\rm BR}(\tau \rightarrow 3\mu) =0$ & ${\rm BR}(\tau \rightarrow 3e) =0$ \\
\hline
 ${\rm G_1}$  &  & $\times$ &  \\
 ${\rm G_2}$  &  & $\times$ & $\times$ \\
 ${\rm G_3}$  & $\times$ & $\times$ &  \\
 ${\rm G_4}$  & $\times$ &  & $\times$ \\
 ${\rm G_5}$  & $\times$ &  & $\times$ \\
 ${\rm G_6}$  & $\times$ & $\times$ & $\times$ \\
 \hline
 ${\rm A_1}$  &  & $\times$ &  \\
 ${\rm A_2}$  &  & $\times$ &  \\
 ${\rm B_1}$  & $\times$ &  &  \\
 ${\rm B_2}$  &  & $\times$ & $\times$ \\
 ${\rm B_3}$  &  &  & $\times$ \\
 ${\rm B_4}$  & $\times$ & $\times$ &  \\
 ${\rm C}$  &  $\times$ & &  $\times$\\
\hline
\end{tabular}
\end{center}
\label{tbl:all_zero}
\end{table}

\begin{table}[t]
\caption{Compatibility of the cases in the one- and two-zero textures scheme with the nonvanishing branching ratios ${\rm BR}(\mu \rightarrow \bar{e}ee)\neq 0$, ${\rm BR}(\tau \rightarrow \bar{\mu}\mu\mu)\neq 0$, and ${\rm BR}(\tau \rightarrow \bar{e}ee)\neq 0$. The symbol $\times$ indicates that the corresponding case should be excluded. }
\begin{center}
\begin{tabular}{|c|c|c|c|}
\hline
 & ${\rm BR}(\mu \rightarrow 3e) \neq 0$ & ${\rm BR}(\tau \rightarrow 3\mu) \neq 0$ & ${\rm BR}(\tau \rightarrow 3e) \neq 0$ \\
\hline
 ${\rm G_1}$  & $\times$ &  & $\times$\\
 ${\rm G_2}$  & $\times$ &  &  \\
 ${\rm G_3}$  &  &  & $\times$ \\
 ${\rm G_4}$  &  & $\times$ &  \\
 ${\rm G_5}$  &  & $\times$ &  \\
 ${\rm G_6}$  &  &  &  \\
 \hline
 ${\rm A_1}$  & $\times$ &  & $\times$ \\
 ${\rm A_2}$  & $\times$  &  & $\times$ \\
 ${\rm B_1}$  &  & $\times$ & $\times$ \\
 ${\rm B_2}$  & $\times$ &  &  \\
 ${\rm B_3}$  & $\times$ & $\times$ &  \\
 ${\rm B_4}$  &  &  & $\times$ \\
 ${\rm C}$  &  & $\times$ &  \\
\hline
\end{tabular}
\end{center}
\label{tbl:all_nonzero}
\end{table}

\subsection{${\rm BR}(\mu\rightarrow 3e,\tau\rightarrow 3\mu, \tau\rightarrow 3e) \neq 0$ case }
In this subsection, we assume that the three lepton flavor violating processes $\mu \rightarrow \bar{e}ee$, $\tau \rightarrow \bar{\mu}\mu\mu$ and $\tau \rightarrow \bar{e}ee$ are all observed experimentally or undoubtedly are predicted theoretically.

In this case, at least the branching ratio ${\rm BR}(\mu \rightarrow \bar{e}ee)$, as well as $M_{ee}$ and $M_{e\mu}$, cannot vanish. The nonvanishing elements $M_{ee}$ and $M_{e\mu}$ ($M_{ee}\neq 0$ and $M_{e\mu}\neq 0$) are inconsistent with the ${\rm G_1}$,${\rm G_2}$,${\rm A_1}$,${\rm A_2}$,${\rm B_2}$, and ${\rm B_3}$ cases in the one- and two-zero textures scheme:
\begin{eqnarray}
&& {\rm G}_1:
\left( 
\begin{array}{*{20}{c}}
0 & \times & \times \\
- & \times & \times \\
- & - & \times \\
\end{array}
\right),
\quad
 {\rm G}_2:
\left( 
\begin{array}{*{20}{c}}
\times & 0 & \times \\
- & \times & \times \\
- & - & \times \\
\end{array}
\right), 
\nonumber \\
&& {\rm A}_1:
\left( 
\begin{array}{*{20}{c}}
0 & 0 & \times  \\
- & \times  & \times \\
- & - & \times \\
\end{array}
\right),
\quad
 {\rm A}_2:
\left( 
\begin{array}{*{20}{c}}
0& \times  &0 \\
- & \times & \times \\
- & - & \times  \\
\end{array}
\right),
\nonumber \\
&& {\rm B}_2:
\left( 
\begin{array}{*{20}{c}}
\times & 0 & \times \\
- & \times & \times \\
- & - & 0 \\
\end{array}
\right),
\quad
{\rm B}_3:
\left( 
\begin{array}{*{20}{c}}
\times & 0 &\times \\
- & 0 & \times \\
- & - & \times \\
\end{array}
\right).
\label{Eq:G1G2B2B3}
\end{eqnarray}
Therefore, the ${\rm G_1}$,${\rm G_2}$,${\rm A_1}$,${\rm A_2}$,${\rm B_2}$ and ${\rm B_3}$ cases in the one- and two-zero textures scheme should be excluded if the lepton flavor violating processes $\mu \rightarrow \bar{e}ee$ are observed experimentally or undoubtedly are predicted theoretically.

Addition to the lepton flavor violating process $\mu \rightarrow \bar{e}ee$, the other two lepton flavor violating processes $\tau \rightarrow \bar{\mu}\mu\mu$ and $\tau \rightarrow \bar{e}ee$ are available for evaluation of the viability of the one- and two-zero textures. Table \ref{tbl:all_nonzero} shows the compatibility of the cases in the one- and two-zero textures schemes with the nonvanishing branching ratios ${\rm BR}(\mu \rightarrow \bar{e}ee)\neq 0$, ${\rm BR}(\tau \rightarrow \bar{\mu}\mu\mu)\neq 0$, and ${\rm BR}(\tau \rightarrow \bar{e}ee)\neq 0$. The symbol $\times$ indicates that the corresponding case should be excluded. 

We conclude that only ${\rm G_6}$ case is viable in one- and two-zero textures of the flavor neutrino mass matrix if the neutrino masses are generated by the type-II seesaw mechanism in triplet Higgs models and the three lepton flavor violating processes $\mu \rightarrow \bar{e}ee$, $\tau \rightarrow \bar{\mu}\mu\mu$ and $\tau \rightarrow \bar{e}ee$ are all observed experimentally or undoubtedly are predicted theoretically.

\subsection{Hybrid cases for $\mu\rightarrow 3e,\tau\rightarrow 3\mu, \tau\rightarrow 3e$\label{subsec:hybrid_cases}}
\begin{table}[t]
\caption{Allowed cases in the one- and two-zero textures scheme for $\mu \rightarrow 3e$, $\tau \rightarrow 3\mu$ and $\tau \rightarrow 3e$. The abbrivation ``NZ" indicates a nonzero value for the branching ratio. }
\begin{center}
\begin{tabular}{cccc}
\hline
${\rm BR}(\mu \rightarrow 3e)$ & ${\rm BR}(\tau \rightarrow 3\mu) $ & ${\rm BR}(\tau \rightarrow 3e) $ & Allowed cases\\
\hline
 0  & 0 & 0 & - \\
 0  & NZ & 0 &  ${\rm G_1}$, ${\rm A_1}$, ${\rm A_2}$\\
 0  & 0 & NZ & ${\rm B_3}$ \\
 0  & NZ & NZ &  ${\rm G_2}$, ${\rm B_4}$\\
 NZ  & 0 & 0 & ${\rm B_1}$ \\
 NZ  & NZ & 0 & ${\rm G_3}$, ${\rm B_4}$ \\
 NZ  & 0 & NZ & ${\rm G_4}$, ${\rm G_5}$, ${\rm C}$ \\
 NZ  & NZ & NZ &  ${\rm G_6}$\\
 \hline
\end{tabular}
\end{center}
\label{tbl:hybrid}
\end{table}

Based on the above discussion, it turned out that if the neutrino masses are generated by the type-II seesaw mechanism in the triplet Higgs models and the three lepton flavor violating processes $\mu \rightarrow \bar{e}ee$, $\tau \rightarrow \bar{\mu}\mu\mu$ and $\tau \rightarrow \bar{e}ee$ are all forbidden, there is no room for one- and two-zero textures. On the other hand, if all three processes exist, only the ${\rm G_6}$ case is viable in one- and two-zero textures. 

If parts of these three  lepton flavor violating processes are allowed such as 
\begin{eqnarray}
&& {\rm BR}(\mu \rightarrow \bar{e}ee) \neq 0, \quad {\rm BR}(\tau \rightarrow \bar{\mu}\mu\mu) =  0, \nonumber \\
&& {\rm BR}(\tau \rightarrow \bar{e}ee) = 0,
\label{Eq:hybrid_ex1}
\end{eqnarray}
other cases of one- and two-zero textures may be allowed. For example, in the case shown in Eq.(\ref{Eq:hybrid_ex1}), the ${\rm G_6}$ case is ruled out and only the ${\rm B_1}$ case is allowed. Similarly, in the cases
\begin{eqnarray}
&& {\rm BR}(\mu \rightarrow \bar{e}ee) = 0, \quad {\rm BR}(\tau \rightarrow \bar{\mu}\mu\mu) \neq  0, \nonumber \\
&& {\rm BR}(\tau \rightarrow \bar{e}ee) = 0,
\label{Eq:hybrid_ex2}
\end{eqnarray}
and
\begin{eqnarray}
&& {\rm BR}(\mu \rightarrow \bar{e}ee) = 0, \quad {\rm BR}(\tau \rightarrow \bar{\mu}\mu\mu) =  0, \nonumber \\
&& {\rm BR}(\tau \rightarrow \bar{e}ee) \neq 0,
\label{Eq:hybrid_ex3}
\end{eqnarray}
the allowed cases of one- and two-zero textures are ${\rm G_1}$, ${\rm A_1}$, ${\rm A_2}$, and $B_3$, respectively. 

Table \ref{tbl:hybrid} shows the allowed cases in the one- and two-zero textures schemes for $\mu \rightarrow 3e$, $\tau \rightarrow 3\mu$ and $\tau \rightarrow 3e$. The abbriviation ``NZ" indicates a nonzero value for the branching ratio.  It is remarkable that the each of  ${\rm G_1}$, ${\rm G_2}$, $\cdots$, ${\rm C}$ cases appears only once in Table \ref{tbl:hybrid}. Therefore, we can predict the allowed combination of nonvanishing branching ratios by the one- and two-zero flavor neutrino mass matrix textures. 

Although whether or not the three lepton flavor violating processes $\mu \rightarrow \bar{e}ee$, $\tau \rightarrow \bar{\mu}\mu\mu$, and $\tau \rightarrow \bar{e}ee$ are forbidden is still undetermined, we can suggest that either
\begin{eqnarray}
{\rm BR}(\mu \rightarrow \bar{e}ee) = {\rm BR}(\tau \rightarrow \bar{\mu}\mu\mu) =  {\rm BR}(\tau \rightarrow \bar{e}ee) = 0,
\nonumber \\
\end{eqnarray}
or
\begin{eqnarray}
{\rm BR}(\mu \rightarrow \bar{e}ee) \neq {\rm BR}(\tau \rightarrow \bar{\mu}\mu\mu) \neq {\rm BR}(\tau \rightarrow \bar{e}ee) \neq 0,
\nonumber \\
\end{eqnarray}
may be the most natural case. Otherwise, the appropriate selection mechanisms for $\ell_m \rightarrow \bar{\ell}_i \ell_j \ell_k$ decay at tree level are required in the models.

We can conclude that if the tiny neutrino masses are generated by the type-II seesaw mechanism, only the ${\rm G_6}$ case may be most natural in one- and two-zero textures schemes. This conclusion becomes more rigid in the next subsection.

\subsection{$\tau \rightarrow \bar{\mu} ee$, $\tau \rightarrow \bar{\mu} e \mu$, $\tau \rightarrow \bar{e} \mu \mu$ and $\tau \rightarrow \bar{e} e \mu$\label{subsec:etc}}

\begin{table*}[t]
\caption{Compatibility of the cases in the one- and two-zero textures schemes with the vanishing branching ratios ${\rm BR}(\tau \rightarrow \bar{\mu} ee)=0$, ${\rm BR}(\tau \rightarrow \bar{\mu} e \mu) =0$, ${\rm BR}(\tau \rightarrow \bar{e} \mu \mu)=0$ and ${\rm BR}(\tau \rightarrow \bar{e} e \mu)=0$. The symbol $\times$ indicates that the corresponding case should be excluded. }
\begin{center}
\begin{tabular}{|c|c|c|c|c|}
\hline
 & ${\rm BR}(\tau \rightarrow \bar{\mu} ee)=0$ & ${\rm BR}(\tau \rightarrow \bar{\mu} e \mu) =0$ & ${\rm BR}(\tau \rightarrow \bar{e} \mu \mu)=0$ & ${\rm BR}(\tau \rightarrow \bar{e} e \mu)=0$ \\
\hline
 ${\rm G_1}$ &   & $\times$ & $\times$ &  $\times$ \\
 ${\rm G_2}$ & $\times$ & & $\times$ &   \\
 ${\rm G_3}$ & $\times$ & $\times$ &  &  \\
 ${\rm G_4}$ & $\times$ & $\times$ & &  $\times$ \\
 ${\rm G_5}$ &  &  & $\times$ &  $\times$ \\
 ${\rm G_6}$ & $\times$ & $\times$ & $\times$ &  $\times$ \\
 \hline
 ${\rm A_1}$ &  &  & $\times$ &  \\
 ${\rm A_2}$ &  & $\times$ &  &   \\
 ${\rm B_1}$ & $\times$ & $\times$ &  &  \\
 ${\rm B_2}$ & $\times$ &  & $\times$ &   \\
 ${\rm B_3}$ & $\times$ &  & &  \\
 ${\rm B_4}$ & $\times$ & $\times$ &  &  \\
 ${\rm C}$  & $\times$ & $\times$ &  &  $\times$ \\
\hline
\end{tabular}
\end{center}
\label{tbl:all_zero2}
\end{table*}

\begin{table*}[t]
\caption{Compatibility of the cases in the one- and two-zero textures schemes with the vanishing branching ratios ${\rm BR}(\tau \rightarrow \bar{\mu} ee)\neq 0$, ${\rm BR}(\tau \rightarrow \bar{\mu} e \mu) \neq 0$, ${\rm BR}(\tau \rightarrow \bar{e} \mu \mu)\neq 0$ and ${\rm BR}(\tau \rightarrow \bar{e} e \mu)\neq 0$. The symbol $\times$ indicates that the corresponding case should be excluded. }
\begin{center}
\begin{tabular}{|c|c|c|c|c|}
\hline
 & ${\rm BR}(\tau \rightarrow \bar{\mu} ee)\neq 0$ & ${\rm BR}(\tau \rightarrow \bar{\mu} e \mu) \neq  0$ & ${\rm BR}(\tau \rightarrow \bar{e} \mu \mu)\neq 0$ & ${\rm BR}(\tau \rightarrow \bar{e} e \mu)\neq  0$ \\
\hline
 ${\rm G_1}$ & $\times$ &  &  &  \\
 ${\rm G_2}$ &  & $\times$ & &  $\times$ \\
 ${\rm G_3}$ &  & & $\times$ &  $\times$ \\
 ${\rm G_4}$ &  &  & $\times$ &\\
 ${\rm G_5}$ & $\times$ & $\times$ &  &  \\
 ${\rm G_6}$ &  &  &  &  \\
 \hline
 ${\rm A_1}$ & $\times$ & $\times$ &  &  $\times$ \\
 ${\rm A_2}$ & $\times$ &  & $\times$ &  $\times$ \\
 ${\rm B_1}$ &  & & $\times$ &  $\times$ \\
 ${\rm B_2}$ &  & $\times$ & &  $\times$ \\
 ${\rm B_3}$ &  & $\times$ & $\times$ &  $\times$ \\
 ${\rm B_4}$ & &  & $\times$ &  $\times$ \\
 ${\rm C}$  &  & & $\times$ &   \\
\hline
\end{tabular}
\end{center}
\label{tbl:all_nonzero2}
\end{table*}

\begin{table*}[t]
\caption{Allowed cases in the one- and two-zero textures schemes for $\tau \rightarrow \bar{\mu} ee$, $\tau \rightarrow \bar{\mu} e \mu$, $\tau \rightarrow \bar{e} \mu \mu$ and $\tau \rightarrow \bar{e} e \mu$. The abbreviation ``NZ" indicates a nonzero value for the branching ratio. }
\begin{center}
\begin{tabular}{ccccc}
\hline
 ${\rm BR}(\tau \rightarrow \bar{\mu} ee)$ & ${\rm BR}(\tau \rightarrow \bar{\mu} e \mu)$ & ${\rm BR}(\tau \rightarrow \bar{e} \mu \mu)$ & ${\rm BR}(\tau \rightarrow \bar{e} e \mu)$  & Allowed cases\\
\hline
 0  & 0 & 0 &0 & - \\
 0  & 0& 0 & NZ & - \\
 0  & 0 & NZ &0 & ${\rm A_1}$ \\
 0  & 0 & NZ & NZ & ${\rm G_5}$ \\
 0  & NZ & 0 &0 & ${\rm A_2}$ \\
 0  & NZ & 0 & NZ & - \\
 0  & NZ & NZ & 0 & - \\
 0  & NZ & NZ & NZ & ${\rm G_1}$ \\
NZ  & 0 & 0 &0 & ${\rm B_3}$ \\
 NZ  & 0& 0 & NZ & - \\
 NZ  & 0 & NZ &0 & ${\rm G_2}$,${\rm B_2}$ \\
 NZ  & 0 & NZ & NZ & - \\
 NZ  & NZ & 0 &0 & ${\rm G_3}$,${\rm B_1}$,${\rm B_4}$ \\
 NZ  & NZ & 0 & NZ & ${\rm G_4}$,${\rm C}$ \\
 NZ  & NZ & NZ & 0 & - \\
 NZ  & NZ & NZ & NZ & ${\rm G_6}$ \\
 \hline
\end{tabular}
\end{center}
\label{tbl:hybrid2}
\end{table*}

In the last subsection, we include only three branching ratios in Eq. (\ref{Eq:BR1}) in our discussion to show our strategy. Now we include the remaining four branching ratios in Eq.(\ref{Eq:BR2}) in our discussion to complete this paper.

According to the same method as in the last subsection, we estimate the compatibility of the cases in the one- and two-zero textures schemes with four branching ratios in Eq. (\ref{Eq:BR2}). The results are shown in Tables \ref{tbl:all_zero2}, \ref{tbl:all_nonzero2} and \ref{tbl:hybrid2}.

Table \ref{tbl:all_zero2} shows the compatibility of the cases in the one- and two-zero textures schemes with the vanishing branching ratios ${\rm BR}(\tau \rightarrow \bar{\mu} ee)=0$, ${\rm BR}(\tau \rightarrow \bar{\mu} e \mu) =0$, ${\rm BR}(\tau \rightarrow \bar{e} \mu \mu)=0$ and ${\rm BR}(\tau \rightarrow \bar{e} e \mu)=0$. We see that all six cases of one-zero textures (${\rm G_1}, {\rm G_2}, \cdots, {\rm C_6}$) and all seven cases of two-zero textures (${\rm A_1}, {\rm A_2}, {\rm B_1},\cdots, {\rm B_4}, {\rm C}$) should be excluded if the neutrino masses are generated by the type-II seesaw mechanism in triplet Higgs models and all four lepton flavor violating processes $\tau \rightarrow \bar{\mu} ee$, $\tau \rightarrow \bar{\mu} e \mu$, $\tau \rightarrow \bar{e} \mu \mu$ and $\tau \rightarrow \bar{e} e \mu$ are explicitly forbidden.

Table \ref{tbl:all_nonzero2} shows the compatibility of the cases in the one- and two-zero textures scheme with the nonvanishing branching ratios ${\rm BR}(\tau \rightarrow \bar{\mu} ee)\neq 0$, ${\rm BR}(\tau \rightarrow \bar{\mu} e \mu) \neq 0$, ${\rm BR}(\tau \rightarrow \bar{e} \mu \mu)\neq 0$ and ${\rm BR}(\tau \rightarrow \bar{e} e \mu)\neq 0$. We see that only the ${\rm G_6}$ case is viable in one- and two-zero textures of the flavor neutrino mass matrix if the neutrino masses are generated by the type-II seesaw mechanism in triplet Higgs models and all four lepton flavor violating processes $\tau \rightarrow \bar{\mu} ee$, $\tau \rightarrow \bar{\mu} e \mu$, $\tau \rightarrow \bar{e} \mu \mu$ and $\tau \rightarrow \bar{e} e \mu$ are observed experimentally or undoubtedly are predicted theoretically.

Table \ref{tbl:hybrid2} shows the allowed cases in the one- and two-zero textures schemes for $\tau \rightarrow \bar{\mu} ee$, $\tau \rightarrow \bar{\mu}, e \mu$, $\tau \rightarrow \bar{e} \mu \mu$ and $\tau \rightarrow \bar{e} e \mu$. The abbreviation ``NZ" indicates a nonzero values for the branching ratio. As in the last subsection, although whether or not the three lepton flavor violating processes $\tau \rightarrow \bar{\mu} ee$, $\tau \rightarrow \bar{\mu} e \mu$, $\tau \rightarrow \bar{e} \mu \mu$ and $\tau \rightarrow \bar{e} e \mu$ are forbidden is still unknown, we can suggest that either
\begin{eqnarray}
{\rm BR}(\tau \rightarrow \bar{\mu} ee) &=& {\rm BR}(\tau \rightarrow \bar{\mu} e \mu) \nonumber \\
 &=&{\rm BR}(\tau \rightarrow \bar{e} \mu \mu) \nonumber \\
 &=&{\rm BR}(\tau \rightarrow \bar{e} e \mu)=0 
\end{eqnarray}
or
\begin{eqnarray}
{\rm BR}(\tau \rightarrow \bar{\mu} ee) &\neq & {\rm BR}(\tau \rightarrow \bar{\mu} e \mu) \nonumber \\
 &\neq &{\rm BR}(\tau \rightarrow \bar{e} \mu \mu) \nonumber \\
 & \neq& {\rm BR}(\tau \rightarrow \bar{e} e \mu)\neq 0
\label{Eq:BRs_not_zero}
\end{eqnarray}
may be the most natural case. 

According to the combined results in the last subsection and this subsection, we conclude that if the tiny neutrino masses are generated by type-II seesaw mechanism, only the ${\rm G_6}$ case may be the most natural in the one- and two-zero textures schemes. This is the main result of this paper.

\subsection{Numerical calculations}
Although the main result of this paper was already obtained in subsection \ref{subsec:etc}, an additional numerical study may be required to improve our discussions. According to the conclusion in subsection \ref{subsec:etc}, only the ${\rm G_6}$ case may be the most natural in the one- and two-zero textures schemes if the tiny neutrino masses have been generated by the type-II seesaw mechanism. In this subsection, we present the phenomenology for the ${\rm G_6}$ case. 

First we give brief reviews of the neutrino mixings, useful relations for the one-zero textures, and observed data from neutrino experiments as a preparation for our numerical calculations. Then we show some predictions for the ${\rm G_6}$ case. 

\

{\bf Neutrino mixings:} 
The flavor neutrino mass matrix $M$ is related to the diagonal neutrino mass matrix
\begin{eqnarray}
M =  U {\rm diag.}(m_1e^{2i\alpha_1},m_2e^{2i\alpha_2} ,m_3) U^T,
\end{eqnarray}
where $m_i$ $(i=1,2,3)$ is a neutrino mass eigenstate and
\begin{eqnarray}
U = \left(
\begin{array}{ccc}
U_{e1}  & U_{e2} & U_{e3} \\
U_{\mu 1}  & U_{\mu 2} & U_{\mu 3} \\
U_{\tau 1}  & U_{\tau 2} & U_{\tau 3} \\
\end{array}
\right),
\end{eqnarray}
with
\begin{eqnarray}
U_{e1} &=& c_{12}c_{13}, \quad U_{e 2} = s_{12}c_{13}, \quad U_{e 3} = s_{13} e^{-i\delta},  \\
U_{\mu 1} &=&- s_{12}c_{23} - c_{12}s_{23}s_{13} e^{i\delta}, \nonumber \\
U_{\mu 2} &=&  c_{12}c_{23} - s_{12}s_{23}s_{13}e^{i\delta}, \quad U_{\mu 3} = s_{23}c_{13}, \nonumber \\
U_{\tau 1} &=& s_{12}s_{23} - c_{12}c_{23}s_{13}e^{i\delta}, \nonumber \\
U_{\tau 2} &=& - c_{12}s_{23} - s_{12}c_{23}s_{13}e^{i\delta}, \quad U_{\tau 3} = c_{23}c_{13},\nonumber 
\end{eqnarray}
denotes the Pontecorvo-Maki-Nakagawa-Sakata (PMNS) mixing matrix \cite{Pontecorvo1957,Pontecorvo1958,Maki1962PTP,PDG}. We use the abbreviations $c_{ij}=\cos\theta_{ij}$ and $s_{ij}=\sin\theta_{ij}$  ($i,j$=1,2,3), where $\theta_{ij}$ is a neutrino mixing angle. The Dirac CP phase is denoted by $\delta$ and the Majorana CP phases are denoted by $\alpha_1$ and $\alpha_2$. In this paper, we assume that the mass matrix of the charged leptons is diagonal and real (some comments for this assumption will be noted in the summary). 

\

{\bf Useful relations for one-zero textures:} 
The requirement of $M_{ij}=0$ for one-zero textures yields
\begin{eqnarray}
A_1 m_1 + A_2 m_2 +A_3 m_3=0,
\end{eqnarray}
where
\begin{eqnarray}
A_1 &=& U_{i1}U_{j1}e^{2i\alpha_1}, \quad A_2 =U_{i2}U_{j2}e^{2i\alpha_2}, \nonumber \\
A_3 &=& U_{i3}U_{j3}.  
\end{eqnarray}
This condition leads to (for examples, see Refs.\cite{Lashin2012PRD,Gautam2018PRD})
\begin{eqnarray}
\frac{m_2}{m_1} = \frac{{\rm Re}(A_1){\rm Im}(A_3) - {\rm Re}(A_3){\rm Im}(A_1) }{{\rm Re}(A_3){\rm Im}(A_2)-{\rm Re}(A_2){\rm Im}(A_3)},
\label{Eq:m2/m1}
\end{eqnarray}
and
\begin{eqnarray}
\frac{m_3}{m_1} = \frac{{\rm Re}(A_2){\rm Im}(A_1) - {\rm Re}(A_1){\rm Im}(A_2) }
{{\rm Re}(A_3){\rm Im}(A_2)-{\rm Re}(A_2){\rm Im}(A_3)}.
\label{Eq:m3/m1}
\end{eqnarray}
The ratio of two squared mass differences is given by
\begin{eqnarray}
\frac{\Delta m_{21}^2}{\left| \Delta m_{31}^2 \right|} = \frac{(m_2/m_1)^2-1}{\left| (m_3/m_1)^2-1\right|},
\label{Eq:m21s_m31s}
\end{eqnarray}
where the squared mass difference is defined by $\Delta m_{ij}^2=m_i^2-m_j^2$. Eqs.(\ref{Eq:m2/m1}), (\ref{Eq:m3/m1}) and (\ref{Eq:m21s_m31s}) are useful when we search the allowed parameter sets under the requirement that $M_{ij}=0$.

\

{\bf Observed data:}
Although the neutrino mass ordering (either the so-called normal mass ordering $m_1 \lesssim m_2<m_3$ or the inverted mass ordering $m_3 < m_1 \lesssim m_2$) has not been determined, a global analysis shows that the preference for normal mass ordering is due mostly to neutrino oscillation measurements \cite{Salas2018PLB, Capozzi2020PRD}. Upcoming experiments for neutrinos will be able to solve this problem \cite{Aartsen2020PRD}. In this paper, we assume the normal mass hierarchical spectrum for the neutrinos. 

A global analysis of current data shows the following best-fit values of the squared mass differences and the mixing angles for the normal mass ordering \cite{Esteban2019JHEP}:
\begin{eqnarray} 
\frac{\Delta m^2_{21}}{10^{-5} {\rm eV}^2} &=& 7.39^{+0.21}_{-0.20} \quad (6.79\rightarrow 8.01), \nonumber \\
\frac{\Delta m^2_{31}}{10^{-3}{\rm eV}^2} &=& 2.528^{+0.029}_{-0.031}\quad (2.436 \rightarrow 2.618), \nonumber \\
\theta_{12}/^\circ &=& 33.82^{+0.78}_{-0.76} \quad (31.61 \rightarrow 36.27), \nonumber \\
\theta_{23}/^\circ &=& 48.6^{+1.0}_{-1.4} \quad (41.1 \rightarrow 51.3), \nonumber \\
\theta_{13}/^\circ &=& 8.60^{+0.13}_{-0.13}\quad (8.22 \rightarrow 8.98), \nonumber \\
\delta/^\circ &=& 221^{+39}_{-28}\quad (144 \rightarrow 357),
\label{Eq:neutrino_observation}
\end{eqnarray}
where $\pm$ signs denote the $1 \sigma$ region and parentheses denote the $3 \sigma$ region. Moreover, the following constraints,
\begin{eqnarray} 
\sum m_i < 0.12 - 0.69 ~{\rm eV},
\end{eqnarray}
from a cosmological observation of cosmic microwave background radiation \cite{Planck2018, Capozzi2020PRD,Giusarma2016PRD,Vagnozzi2017PRD,Giusarma2018PRD} as well as 
\begin{eqnarray} 
|M_{ee}|<0.066 - 0.155 ~{\rm eV},
\end{eqnarray}
from the neutrinoless double beta decay experiments \cite{GERDA2019Science,Capozzi2020PRD} are obtained.

\

{\bf Phenomenology for ${\rm G_6}$ case:} 
Now we make some predictions for the ${\rm G_6}$ case by using numerical calculations.

In our numerical calculation, we require that the squared mass differences $\Delta m^2_{ij}$, mixing angles $\theta_{ij}$, and the Dirac CP violating phase $\delta$ are varied within the $3 \sigma$ experimental ranges, the Majorana CP violating phases $\alpha_1$ and $\alpha_2$ are varied within their full possible ranges and the lightest neutrino mass is varied within $0.01 - 0.1$ eV. We also require that the constraints $|M_{ee}| < 0.155$ eV and $\sum m_i < 0.241$ eV (TT, TE, EE+LowE+lensing \cite{Planck2018,Singh2020EPL}) are satisfied. As predictions for the one-zero textures, we estimate the ratios
\begin{eqnarray}
\label{Eq:R}
R_1&=&\frac{{\rm BR}(\tau \rightarrow \bar{\mu} \mu \mu)}{{\rm BR}(\tau \rightarrow \bar{e} e e)} = 
\frac{ |M_{\mu \tau}|^2 |M_{\mu\mu}|^2}{ |M_{e \tau}|^2 |M_{ee}|^2},  \\
R_2&=&\frac{{\rm BR}(\mu \rightarrow \bar{e} e e)}{{\rm BR}(\tau \rightarrow \bar{e} e e)} = 
\frac{ |M_{e \mu}|^2}{ |M_{e \tau}|^2}\frac{{\rm BR}(\mu \rightarrow e \bar{\nu} \nu)}{{\rm BR}(\tau \rightarrow \mu \bar{\nu} \nu)}, \nonumber \\
R_3&=&\frac{{\rm BR}(\mu \rightarrow \bar{e} e e)}{{\rm BR}(\tau \rightarrow \bar{\mu} \mu \mu)} = 
\frac{ |M_{e \mu}|^2 |M_{ee}|^2}{ |M_{\mu \tau}|^2 |M_{\mu\mu}|^2}\frac{{\rm BR}(\mu \rightarrow e \bar{\nu} \nu)}{{\rm BR}(\tau \rightarrow \mu \bar{\nu} \nu)}, \nonumber
\end{eqnarray}
where ${\rm BR}(\mu \rightarrow e \bar{\nu} \nu) \simeq 100 \%$ and ${\rm BR}(\tau \rightarrow \bar{\mu} \nu \nu) \simeq 17.39\%$ \cite{PDG}.

We show an example of the results of our numerical calculations for the ${\rm G_6}$ case. A point set
\begin{eqnarray} 
(\theta_{12}, \theta_{23},\theta_{13}, \delta) &=& (33.82^\circ, 48.6^\circ, 8.60^\circ,221^\circ), \nonumber \\
(\alpha_1, \alpha_2) &=& (90.03^\circ, 89.2^\circ), \nonumber \\
m_1&=& 0.0580 {\rm eV}
\end{eqnarray}
yields the following neutrino flavor masses
\begin{eqnarray}
M_{ee} &=& -0.0567 - 0.00125i, \nonumber \\ 
M_{e\mu}&=& -0.0115 + 0.00191i, \nonumber \\ 
M_{e\tau}&=& -0.00979 + 0.000647i, \nonumber \\ 
M_{\mu\mu}&=& 0.0165 - 0.000176i, \nonumber \\ 
M_{\mu\tau}&=& 0.0661 - 0.00120 i, \nonumber \\ 
M_{\tau\tau}&=& 0, 
\end{eqnarray}
%
as well as 
\begin{eqnarray}
&& m_2= 0.0586~{\rm eV}, \quad m_3= 0.0768 ~ {\rm eV}, \nonumber \\
&& \Delta m^2_{21} = 6.95 \times 10^{-5} ~{\rm eV}^2,\quad \Delta m^2_{31}= 2.53 \times 10^{-3}~ {\rm eV}^2, \nonumber \\
&& \sum m_i = 0.193~{\rm  eV}, \quad  \vert M_{ee} \vert= 0.0567 ~ {\rm  eV}. 
\end{eqnarray}
These results are consistent with observations. The predicted ratios [Eq.(\ref{Eq:R})] are
\begin{eqnarray}
 (R_1, R_2, R_3) = (3.85, 8.12, 2.11).
\end{eqnarray}

\begin{figure}[t]
\begin{center}
\includegraphics{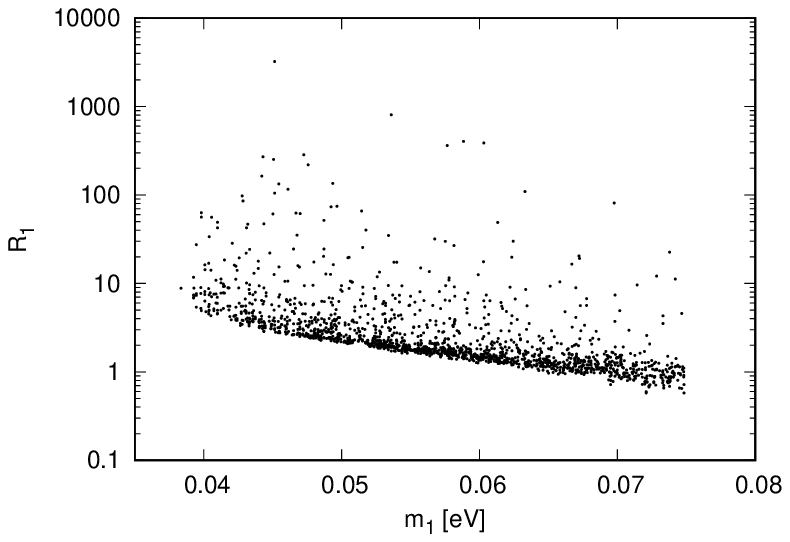}\\
\includegraphics{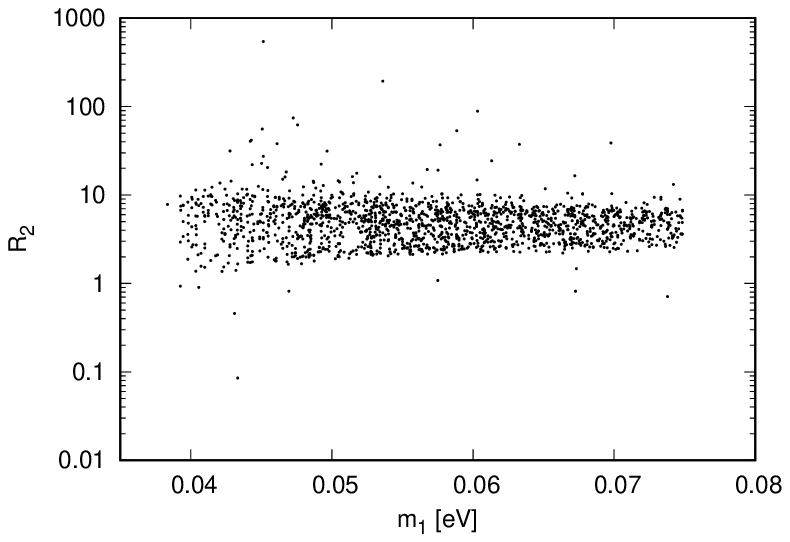}\\
\includegraphics{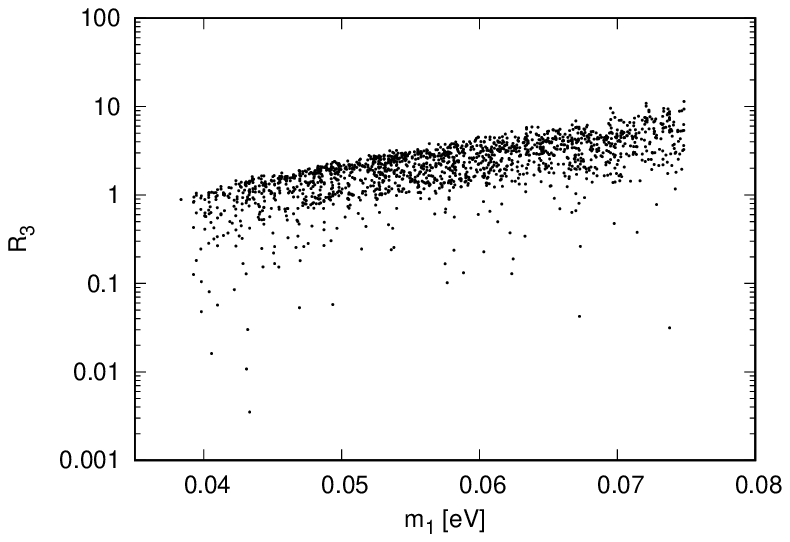}
\caption{Predictions for $R_1$, $R_2$, and $R_3$ for the lightest neutrino mass $m_1$ for the ${\rm G_6}$ case within the type-II seesaw generation of the neutrino masses in triplet Higgs models.}
\label{fig:R_m1}
\end{center}
\end{figure}

Figure \ref{fig:R_m1} shows that the predictions for $R_1$, $R_2$ and $R_3$ for the lightest neutrino mass $m_1$ for the ${\rm G_6}$ case. Currently, we have only the upper limits of ${\rm BR}(\tau \rightarrow \bar{\mu} \mu \mu) < 2.1 \times 10^{-8}$, ${\rm BR}(\tau \rightarrow \bar{e} e e) < 2.7 \times 10^{-8}$ and ${\rm BR}(\mu \rightarrow \bar{e} e e) < 1.0 \times 10^{-12}$ from observations \cite{PDG}. If these branching ratios are determined in the future experiments, 
\begin{eqnarray}
R_1 &\simeq& 0.6 - 3234,\nonumber \\
R_2 &\simeq& 0.07 - 542, \nonumber \\
R_3 &\simeq& 0.004 - 11, 
\end{eqnarray}
support is given to the ${\rm G_6}$ case within the type-II seesaw generation of the neutrino masses in triplet Higgs models.

Finally, we would like to mention the very recently reported tension between NOvA and T2K in the measurement of $\delta$ and $\sin^2\theta_{23}$ for the normal mass ordering of neutrinos \cite{Himmel2020,Dunne2020}. Both experiments favor the upper octant of $\theta_{23}$; however, the NOvA data show $\delta < \pi$, which is contrary to the T2K result $\delta >\pi$. In this paper, until now, we have used the data from the global analysis shown in Eq.(\ref{Eq:neutrino_observation}) for our numerical calculations. The $3\sigma$ data in this global analysis, the upper octant of $\theta_{23}$ and $\delta > \pi$ are roughly favored. 

\begin{figure}[t]
\begin{center}
\includegraphics{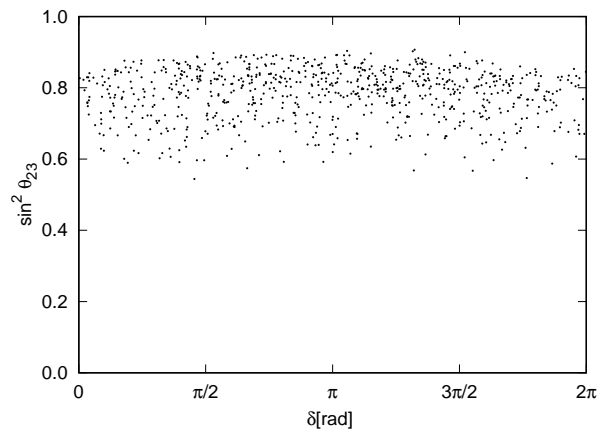}
\caption{Allowed parameter space of $\theta_{23}$ and $\delta$ for the ${\rm G_6}$ case with the normal mass ordering of neutrinos.}
\label{fig:sin2theta23_delta}
\end{center}
\end{figure}

If the mixing angle $\theta_{23}$ and the CP phase $\delta$ are varied within their full range (e.g., $0^\circ \le \theta_{23} \le 90^\circ$ and $0^\circ \le \delta \le 360^\circ$) to try to obtain insight into the tension between NOvA and T2K; unfortunately, we could not obtain a significant prediction for this tension. Figure \ref{fig:sin2theta23_delta} shows the allowed parameter space of $\sin^2\theta_{23}$ and $\delta$ for the ${\rm G_6}$ case with the normal mass ordering of neutrinos. The upper octant of $\theta_{23}$ is favored in the ${\rm G_6}$ case. It is consistent with NOvA and T2K observations; however, the broad region is allowed for the Dirac CP phase $\delta$ in the ${\rm G_6}$ case.

\section{\label{sec:summary}Summary}

One- and two-zero textures for the flavor neutrino mass matrix have been successful in explaining mixing in the neutrino sector. In this paper, we have shown that all cases of one- and two-zero textures are excluded if the tiny neutrino masses are generated by the type-II seesaw mechanism in triplet Higgs models and the three lepton flavor violating processes $\mu \rightarrow \bar{e}ee$, $\tau \rightarrow \bar{\mu}\mu\mu$ and $\tau \rightarrow \bar{e}ee$ are all explicitly forbidden experimentally or theoretically. We have also shown that if all three of these lepton flavor violating processes exist, only the ${\rm G_6}$ case is viable within the one- and two-zero textures. 

Even if parts of these three lepton flavor violating processes are allowed, such as ${\rm BR}(\mu \rightarrow 3e) \neq 0$, ${\rm BR}(\tau \rightarrow 3\mu) =  0$ and ${\rm BR}(\tau \rightarrow 3e) = 0$, we can suggest that the most natural case is either ${\rm BR}(\mu \rightarrow 3e) = {\rm BR}(\tau \rightarrow 3\mu) = {\rm BR}(\tau \rightarrow 3e) = 0$ or ${\rm BR}(\mu \rightarrow 3e) \neq {\rm BR}(\tau \rightarrow 3\mu) \neq {\rm BR}(\tau \rightarrow 3e) \neq 0$. Otherwise, the appropriate selection mechanisms for $\ell_m \rightarrow \bar{\ell}_i \ell_j \ell_k$ decay at tree level are required in the models. Therefore we have concluded that if the tiny neutrino masses are generated by the type-II seesaw mechanism in the triplet Higgs models, only the ${\rm G_6}$ case may be most natural in one- and two-zero textures schemes. We have also shown that this conclusion becomes more rigid by including four other lepton flavor violating processes, $\tau \rightarrow \bar{\mu} ee$, $\tau \rightarrow \bar{\mu} e \mu$, $\tau \rightarrow \bar{e} \mu \mu$ and $\tau \rightarrow \bar{e} e \mu$ in our discussions.

Moreover, some predictions for the ${\rm G_6}$ case have been made. The ratios $R_1={\rm BR}(\tau \rightarrow \bar{\mu} \mu \mu)/{\rm BR}(\tau \rightarrow \bar{e} e e)$, $R_2={\rm BR}(\mu \rightarrow \bar{e} e e)/{\rm BR}(\tau \rightarrow \bar{e} e e)$ and $R_3={\rm BR}(\mu \rightarrow \bar{\mu} \mu \mu)/{\rm BR}(\tau \rightarrow \bar{e} e e)$  should be $R_1 \simeq  0.6 - 3234$, $R_2 \simeq 0.07 - 542$ and $R_3 \simeq 0.004 - 11$, respectively, for the ${\rm G_6}$ case within the type-II seesaw generation of the neutrino masses in the triplet Higgs models. In light of recent tension between NOvA and T2K in the measurement of $\delta$ and $\sin^2\theta_{23}$ for the normal mass ordering of neutrinos, we have estimated the allowed parameter space of $\theta_{23}$ and $\delta$ for the ${\rm G_6}$ case; however, we have no significant prediction for this tension. 

Finally, we would like to mention the role of the charged lepton mixings. In general, the lepton mixing (PMNS) matrix $U$ is obtained as \cite{Hochmuth2007PLB}
\begin{eqnarray}
U=U_\ell^\dag U_\nu,
\end{eqnarray}
where
\begin{eqnarray}
U_{\ell(\nu)} = \left(
\begin{array}{ccc}
U_{11}^{\ell(\nu)}   & U_{12}^{\ell(\nu)} & U_{13}^{\ell(\nu)} \\
U_{21}^{\ell(\nu)}  & U_{22}^{\ell(\nu)} & U_{23}^{\ell(\nu)} \\
U_{31}^{\ell(\nu)}  & U_{32}^{\ell(\nu)} & U_{33}^{\ell(\nu)} \\
\end{array}
\right),
\end{eqnarray}
with
\begin{eqnarray}
U_{11}^{\ell(\nu)} &=& c_{12}^{\ell(\nu)}c_{13}^{\ell(\nu)}, \quad U_{12}^{\ell(\nu)} = s_{12}^{\ell(\nu)}c_{13}^{\ell(\nu)}, \nonumber \\
U_{13}^{\ell(\nu)} &=& s_{13}^{\ell(\nu)} e^{-i\delta_{\ell(\nu)}},  \nonumber \\
U_{21}^{\ell(\nu)} &=&- s_{12}^{\ell(\nu)}c_{23}^{\ell(\nu)} - c_{12}^{\ell(\nu)}s_{23}^{\ell(\nu)}s_{13}^{\ell(\nu)} e^{i\delta_{\ell(\nu)}}, \nonumber \\
U_{22}^{\ell(\nu)} &=&  c_{12}^{\ell(\nu)}c_{23}^{\ell(\nu)} - s_{12}^{\ell(\nu)}s_{23}^{\ell(\nu)}s_{13}^{\ell(\nu)}e^{i\delta_{\ell(\nu)}}, \nonumber \\
U_{23}^{\ell(\nu)} &=& s_{23}^{\ell(\nu)}c_{13}^{\ell(\nu)}, \nonumber \\
U_{31}^{\ell(\nu)} &=& s_{12}^{\ell(\nu)}s_{23}^{\ell(\nu)} - c_{12}^{\ell(\nu)}c_{23}^{\ell(\nu)}s_{13}^{\ell(\nu)}e^{i\delta_{\ell(\nu)}}, \nonumber \\
U_{32}^{\ell(\nu)} &=& - c_{12}^{\ell(\nu)}s_{23}^{\ell(\nu)} - s_{12}^{\ell(\nu)}c_{23}^{\ell(\nu)}s_{13}^{\ell(\nu)}e^{i\delta_{\ell(\nu)}}, \nonumber \\
U_{33}^{\ell(\nu)} &=& c_{23}^{\ell(\nu)}c_{13}^{\ell(\nu)}.
\end{eqnarray}
We used the abbreviations $c_{ij}^{\ell(\nu)}=\cos\theta_{ij}^{\ell(\nu)}$ and $s_{ij}^{\ell(\nu)}=\sin\theta_{ij}^{\ell(\nu)}$  ($i,j$=1,2,3) where $\theta_{ij}^{\ell(\nu)}$ is a mixing angle in the charged lepton (neutrino) sector. $\delta_{\ell(\nu)}$ denotes the CP violating phase in the charged lepton (neutrino) sector.

As the observables, the sine and cosine of the three mixing angles of the PMNS matrix $U$ are given by
\begin{eqnarray}
&& s_{12}^2=\frac{|U_{e2}|^2}{1-|U_{e3}|^2}, \quad s_{23}^2=\frac{|U_{\mu 3}|^2}{1-|U_{e3}|^2}, \quad s_{13}^2=|U_{e3}|^2, \nonumber \\
&& c_{12}^2= \frac{|U_{e1}|^2}{1-|U_{e3}|^2}, \quad c_{23}^2 = \frac{|U_{\tau 3}|^2}{1-|U_{e3}|^2}.
\end{eqnarray}
For example, since we obtain the following relations
\begin{eqnarray}
U_{e3} &=& U_{11}^{\ell *} U_{13}^{\nu} + U_{21}^{\ell *} U_{23}^{\nu} + U_{31}^{\ell *} U_{33}^{\nu},  \nonumber \\
U_{\mu 3} &=& U_{12}^{\ell *} U_{13}^{\nu} + U_{22}^{\ell *} U_{23}^{\nu} + U_{32}^{\ell *} U_{33}^{\nu},
\end{eqnarray}
the lepton mixing angle $\theta_{23}$ depends not only on the mixing angles in the neutrino sector but also on the mixing angles in the charged lepton sector. Thus, the predicted $\sin^2 \theta_{23}$ in the Figure \ref{fig:sin2theta23_delta} should be modified if the charged lepton mixing matrix $U_\ell$ is no longer the identity matrix. The detail of this modification depends on the models of the charged lepton mixings. 

In this paper, we have assumed that the mass matrix of the charged leptons is diagonal and real.  In this case, we obtain $U=U_\nu$ for $U_\ell=I$ where $I$ denotes the identity matrix; however, once condition {\bf C3} or Eq.(\ref{Eq:BRs_not_zero}) is allowed, then the charged lepton mass matrix is no longer diagonal and $U_\ell$ is no longer an identity matrix, and hence $U=U_\ell^\dag U_\nu$.

Our assumption, $U=U_\nu$, should be interpreted as $U \sim U_\nu$ with $U_\ell \sim I$. In this case, the contribution coming from charged lepton sector should be negligible, which is possible only if the branching ratios of the lepton flavor violating processes are far below the experimental limits; this nealy leads to the  {\bf C2} condition. Thus, the abbreviation ``NZ" in Tables \ref{tbl:hybrid} and \ref{tbl:hybrid2} indicates nonvanishing but tiny values. 

If the branching ratios have measurable magnitudes for experiments in the near future, a more significant contribution of the charged leptons to the lepton mixing matrix may be necessary. We would like to discuss the details of this topic in a separate work in the future.

\vspace{3mm}






\end{document}